\documentclass[twocolumn,showpacs,preprintnumbers,amsmath,amssymb,supersrcriptaddress]{revtex4-1}

\usepackage{graphicx}
\usepackage{dcolumn}
\usepackage{bm}

\begin{document}

\preprint{APS/XX0000}

\title{Neutrino-induced neutral- and charged-current reactions on $^{40}$Ar}

\author{Toshio Suzuki$^{1,2,3}$}
\email{suzuki.toshio@nihon-u.ac.jp}

\author{Noritaka Shimizu$^{4}$}

\affiliation{$^{1}$Department of Physics,
College of Humanities and Sciences, Nihon University     
Sakurajosui 3-25-40, Setagaya-ku, Tokyo 156-8550, Japan
}
\affiliation{
$^{2}$NAT Research Center, 3129-45 Hibara Muramatsu, Tokai-mura, Naka-gun, Ibaraki 319-1112, Japan}
\affiliation{
$^{3}$School of Physics, Beihang University, 37 Xueyuan Road, Haidian-qu, Beijing 100083, People's Republic of China}

\affiliation{$^{4}$Center for Computational Sciences, University of Tsukuba, 1-1-1 Tennodai, Tsukuba, Ibaraki 305-8577 Japan} 

%
\date{\today}

\begin{abstract}
Neutrino-induced reactions on $^{40}$Ar are investigated by shell model for Gamow-Teller transitions and random-phase-approximation (RPA) for forbidden transitions.
For the 1$^{+}$ multipole, an effective interaction in $sd$-$pf$ shell obtained by the extended Kuo-Krenciglowa (EKK) method from chiral interactions is used to study $B(GT)$, charged-current reaction $^{40}$Ar ($\nu_e$, $e^{-}$) $^{40}$K, $B(M1)$ in $^{40}$Ar and neutral-current reaction $^{40}$Ar ($\nu$, $\nu$') $^{40}$Ar.
A considerable quenching for spin modes is found in the analysis of $B(M1)$, and this quenching is taken into account for the evaluation of the cross sections of the neutral-current reaction.
The sensitive dependence of the reaction cross sections on the quenching of the axial-vector coupling constant, $g_A$, is pointed out. 
\end{abstract}

\pacs{25.30.Pt, 21.60.Cs, 21.60.Jz, 27.40.+z}
\maketitle


\def\be{\begin{equation}}
\def\ee{\end{equation}}
\def\bea{\begin{eqnarray}}
\def\eea{\end{eqnarray}}
\def\br{\bf r}


    



\section{Introduction}
The study of low-energy neutrinos and neutrino-nucleus interactions is important for unveiling the properties of neutrinos such as mass hierarchy and CP-violating phase, which are still open problems, as well as physics beyond the standard model such as neutrino magnetic moment, non-standard interactions and sterile neutrinos \cite{Ankowski}.
Detection of supernova neutrinos is crucial to study supernova dynamics, neutrino oscillations in matter and nucleosynthesis \cite{Balasi}. 
Liquid Ar detectors such as Liquid Argon Time Projection Chambers (LATPC) \cite{LAR} are important tools for the study of neutrino properties by detection of supernova and decay-at-rest (DAR) neutrinos. 
The DAR neutrinos are now available at the Spallation Neutron Source at Oak Ridge National Laboratory (ORNL) \cite{ORNL}.
Detection of supernova neutrinos is planned at Super-Kamiokande \cite{superK}, Hyper-Kamiokande \cite{hyperK}, the Deep Underground Neutrino Experiment (DUNE) \cite{dune} and the Jiangmen Underground Neutrino Observatory (JUNO) \cite{juno}.

Here, neutrino-induced reactions on $^{40}$Ar are studied for neutrino energies at $E_{\nu} \leq$ 100 MeV by a hybrid model \cite{Langanke,SH2013}, where Gamow-Teller (GT) transitions are treated by shell model while random-phase approximation (RPA) is employed for forbidden transitions. 
The multipole expansion method of Walecka \cite{Walecka} is used for the evaluations of neutrino-induced reaction cross sections. 
The GT part of the charged-current reaction $^{40}$Ar ($\nu_e$, $e^{-}$) $^{40}$K was investigated with the use of a shell-model Hamiltonian for $sd$-$pf$ shell \cite{SH2013}, where the $sd$-$pf$ cross-shell is taken to be the monopole-based universal interaction (VMU) \cite{OSHU2010}.
The VMU consists of tensor components of $\pi$+$\rho$ meson exchanges and central components with one-range Gaussian form.
The monopole terms of the tensor interaction have a general sign rule: attractive for $j_{>}$ = $\ell$+1/2 and $j_{<}$ = $\ell$-1/2 orbits and repulsive for $j_{>}$-$j_{>}$ and $j_{<}$-$j_{<}$ orbits \cite{OSFG}.
Monopole terms of microscopic G-matrix and good phenomenological interactions such as SDPF-M \cite{Utsuno} and GXPF1A \cite{Honma} have characteristic orbit dependences, that is, a kinked structure consistent with this general rule, and this feature can be attributed to the tensor components of the interactions \cite{OSHU2010}. 
The important roles of the tensor interaction are thus universal in effective interactions, and the use of the VMU for the cross-shell part is based on the general features of the monopole terms of the tensor interaction shown above.
  
The experimental GT strength obtained by (p, n) reactions \cite{Bhat} was found to be well reproduced by these shell-model studies \cite{SH2013}.
Shell-model calculations can take into account more correlation effects compared to the RPA methods, while the cross sections were also obtained by QRPA \cite{Cheoun2011} and CRPA calculations \cite{Jacho}.         
Here, we use another shell-model Hamiltonian for $sd$-$pf$ shell recently obtained by the extended Kuo-Krenciglowa (EKK) method from effective chiral interactions \cite{TO2017}.
The interaction proves to be quite successful in explaining the spectroscopic properties and driplines of isotopes with charge number $Z$ = 9-12 \cite{TO2017,TO2020}. 
The GT strength and cross sections for $^{40}$Ar ($\nu_e$, $e^{-}$) $^{40}$K evaluated by the new effective interaction with the EKK method are discussed in Sect. 2.
In Sect. 3, magnetic dipole (M1) transition strength and neutral-current reaction cross sections for $^{40}$Ar are evaluated by the effective interaction obtained by the EKK method for the 1$^{+}$ multipole and by RPA for forbidden transitions.
The quenching for the axial-vector coupling constant $g_A$ is discussed.
The summary is given in Sect. 4.


\begin{figure*}[tbh]
\hspace{-5mm}
\includegraphics[scale=0.49]{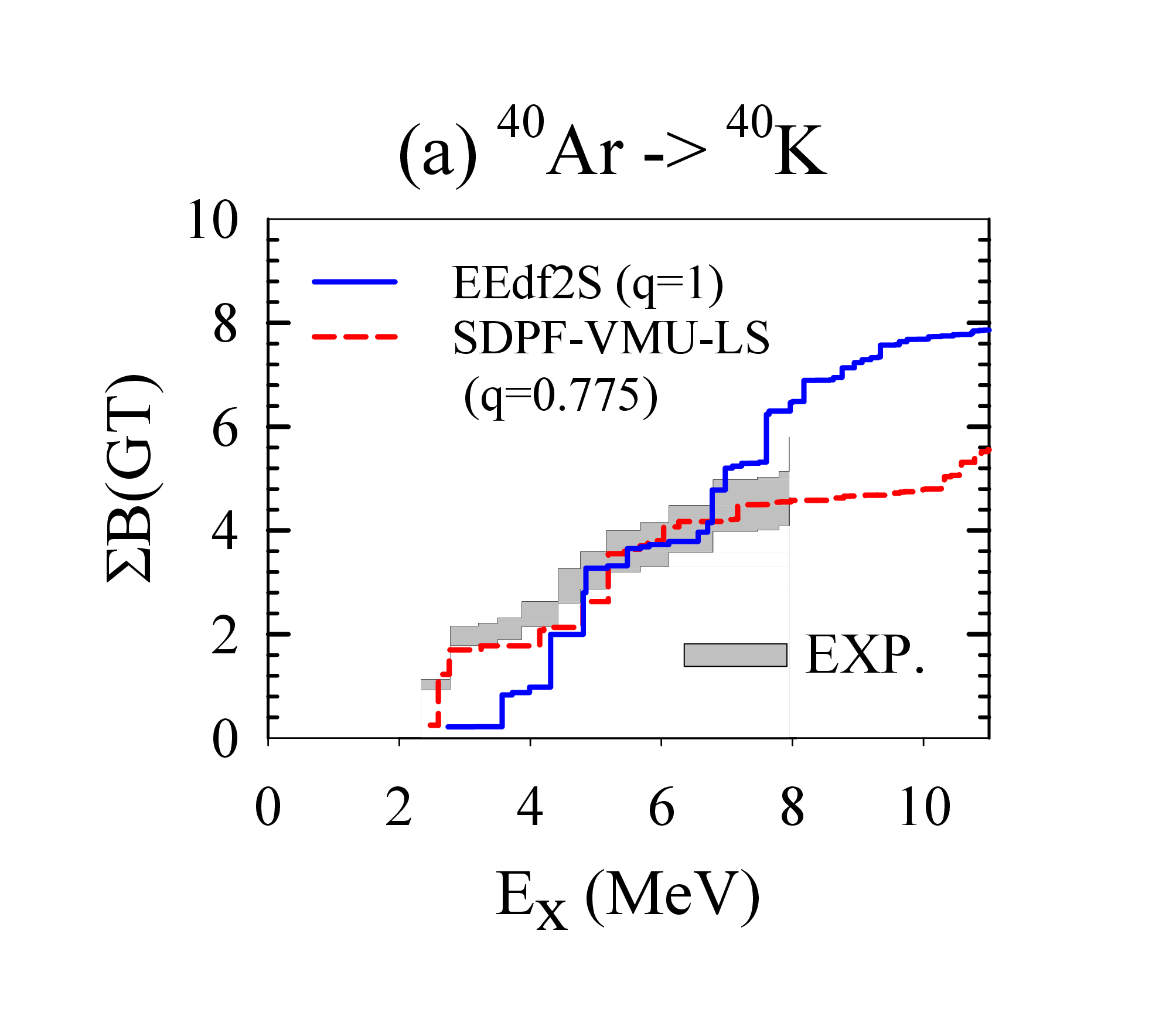}
\hspace{-15mm}
\includegraphics[scale=0.44]{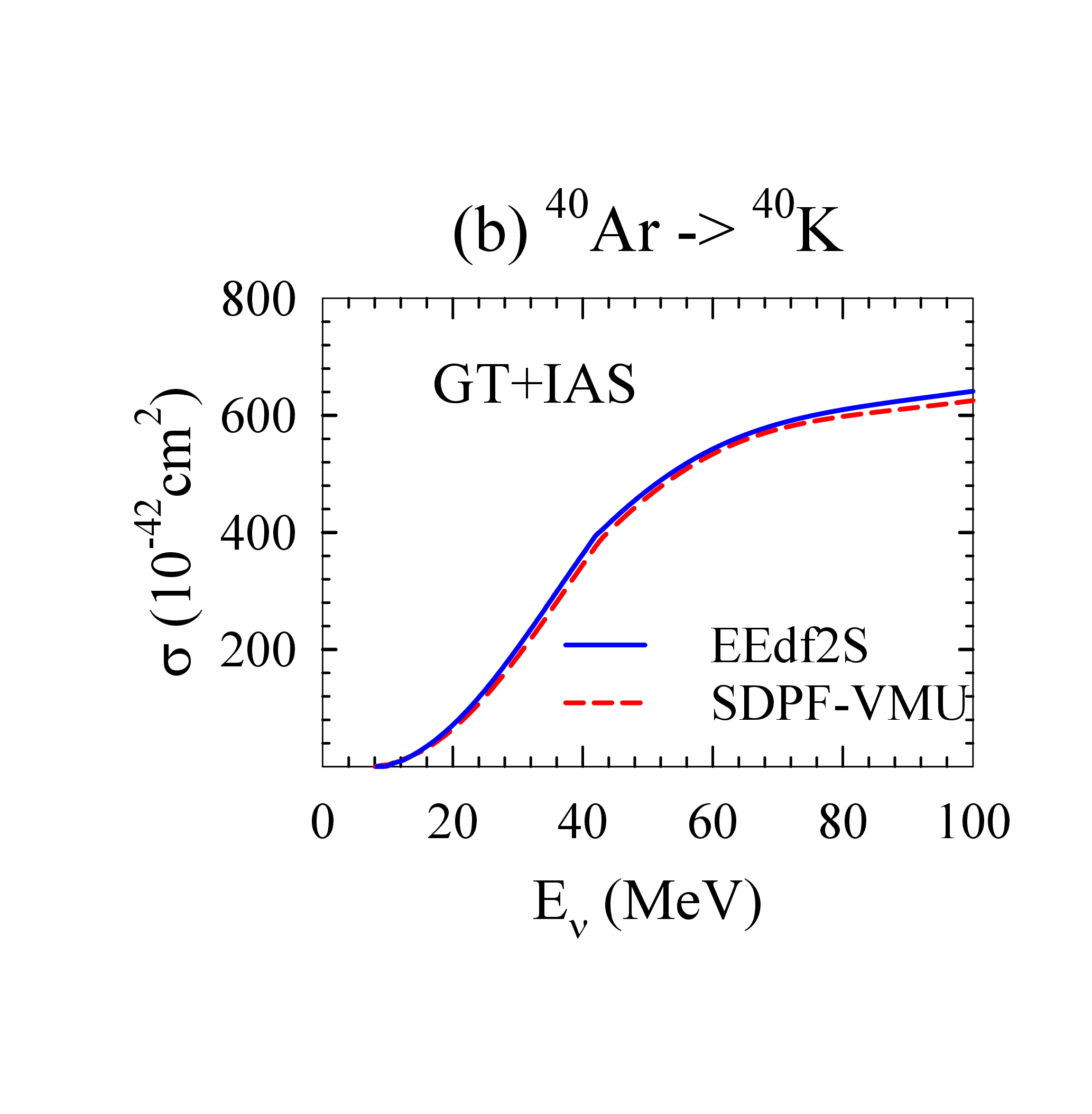}
\vspace{-5mm}
\caption{
(Color online) (a) Cumulative sum of the GT strength for $^{40}$Ar $\rightarrow$ $^{40}$K up to excitation energies of $^{40}$K, $E_x$ , obtained by the shell-model calculation with the use of EEdf2S and SDPF-VMU-LS \cite{SH2013} interactions. 
The experimental data \cite{Bhat} are shown by the shaded area.
(b) Calculated reaction cross sections for $^{40}$Ar ($\nu_e$, $e^{-}$) $^{40}$K.  Contributions from the GT and isobaric analog (IA) transitions obtained by the shell-model calculations with the use of EEdf2S and SDPF-VMU-LS are shown. 
\label{fig:fig1}}
\end{figure*}

\section{Charged-current reactions}

\begin{figure*}[tbh]
\hspace{-5mm}
\includegraphics[scale=0.44]{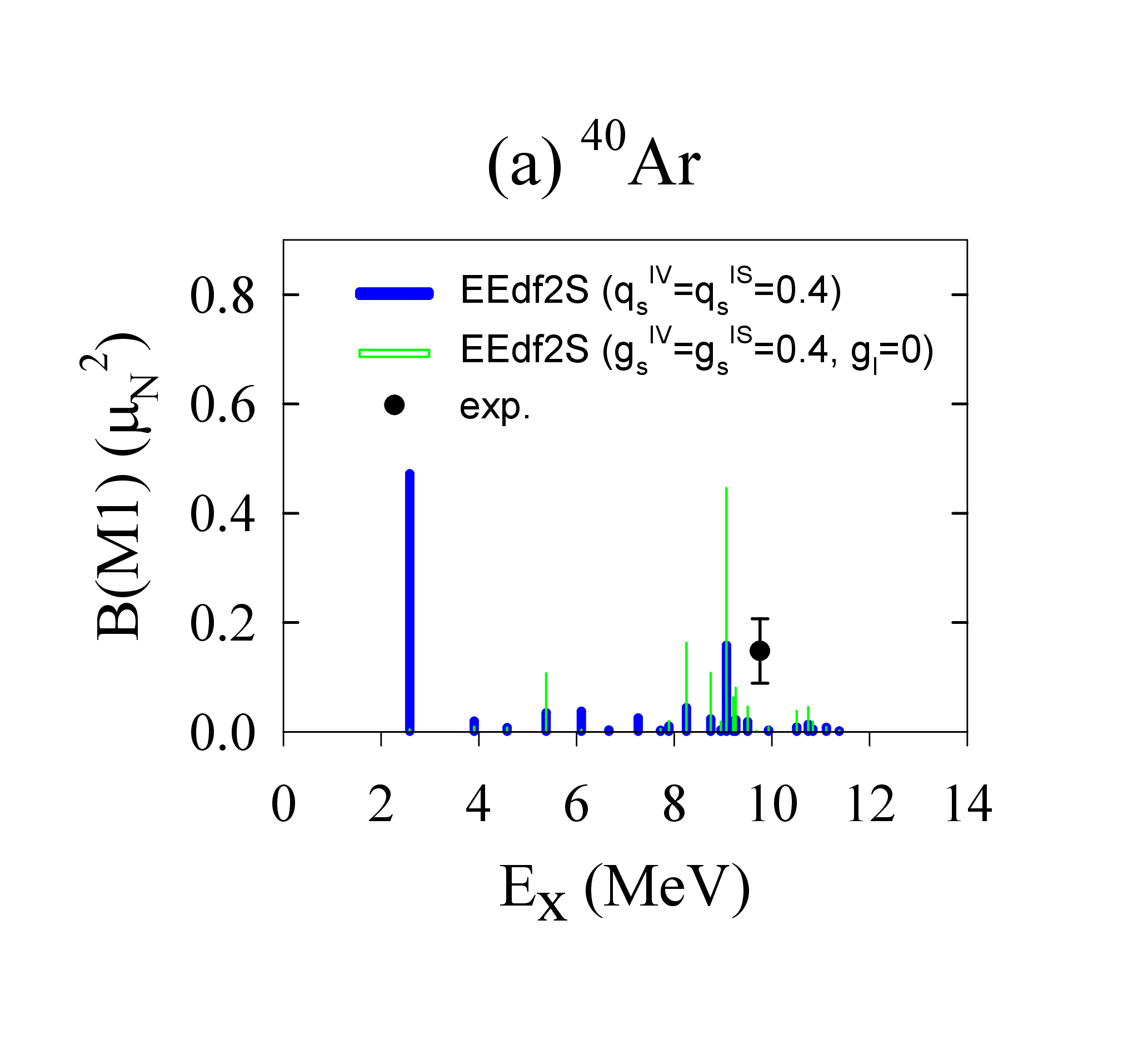}
\hspace{-5mm}
\includegraphics[scale=0.44]{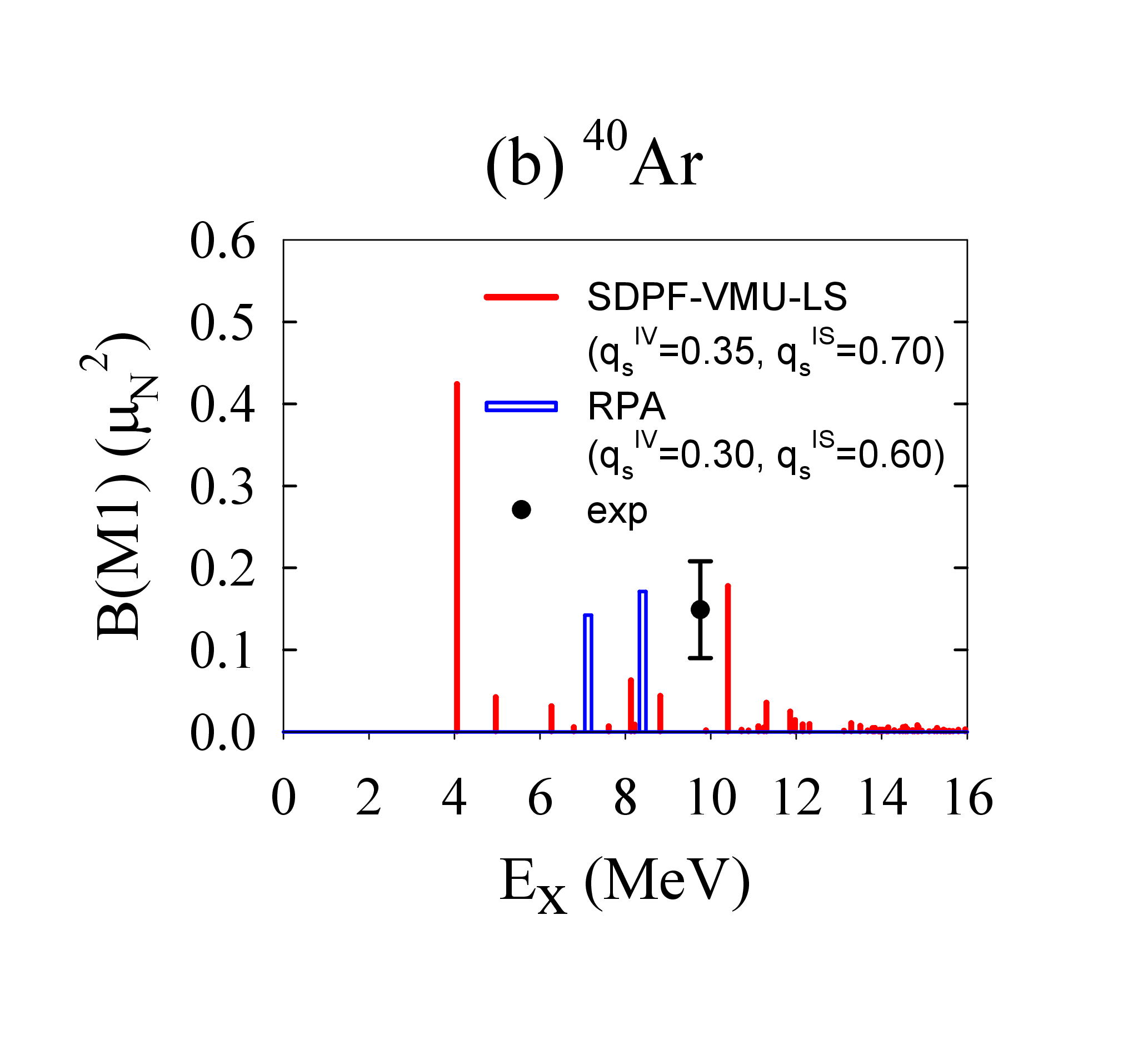}
\vspace{-10mm}
\caption{(Color online)
(a) Calculated $B(M1)$ values in $^{40}$Ar obtained by shell-model calculations with the EEdf2S interaction.
The quenching factors for the spin $g$-factors are taken to be $q_s^{IV}$ =$q_s^{IS}$ =0.4.
Thin hollow histograms are obtained with $g_{\ell}$ =0.  Experimental data \cite{Li2006} is also denoted.
(b) The same as in (a) for the shell-model calculations with the SDPF-VMU-LS \cite{SH2013} interaction with the quenching factors $q_s^{IV}$=0.35 and $q_s^{IS}$=0.70 (red filled histograms), and for RPA calculations with the SGII \cite{SG2} with the quenching factors $q_s^{IV}$ =0.30 and $q_s^{IS}$ =0.60 (blue hollow histograms).
\label{fig:fig2}}
\end{figure*}

In the conventional G-matrix method \cite{Kuo}, effective interactions are obtained within one-major shell, but the method is not applicable to the case of two-major shells because of the divergence problem in the Q-box expansion \cite{TTJO2014}.
The VMU, therefore, is often used for the cross-shell part of the interaction and proves to be successful in many cases, for example, in SDPF-MU \cite{SDPFMU} for  $sd$-$pf$ shell, YSOX \cite{YS2012} for $p$-$sd$ shell and effective interactions in the Pb region \cite{Yuan2022}.
The $sd$-$pf$ cross-shell part of the effective interaction used for $^{40}$Ar, referred to as SDPF-VMU-LS \cite{SH2013}, includes also the two-body spin-orbit interaction from meson-exchanges, and $sd$- and $pf$- shell parts are those of phenomenological SDPF-M \cite{Utsuno} and GXPF1A \cite{Honma} interactions, respectively.       
  
Recently, the EKK method was applied to the $sd$-$pf$ shell starting from the chiral N$^{3}$LO interaction \cite{Machleidt} including up to the third-order Q-box expansions \cite{TO2017}. 
The effective interaction with the additional Fujita-Miyazawa three-nucleon (3N) interaction \cite{FM}, referred to as EEdf1 \cite{OGSSU}, is found to explain the energy spectra, electric quadrupole (E2) transition strengths, and driplines of the F, Ne, Na and Mg isotopes quite well \cite{TO2020}.
In the present work, we use the effective interaction with the chiral N$^{2}$LO 3N interaction \cite{Gazit} instead of the Fujita-Miyazawa force. 
Density-dependent nucleon-nucleon (NN) interaction is first derived from the 3N interaction by folding over the third nucleon in the Fermi sea \cite{Kohno}, and then the effective NN interaction is obtained by having the density dependence integrated out with the normal density \cite{TO2020}.
The new interaction, which will be referred to as EEdf2S hereafter, can also explain the spectroscopic properties of the neutron-rich isotopes as well. 

The GT strength in $^{40}$Ar is evaluated by shell-model calculations with the use of the EEdf2S in $sd^{-2}pf^{2}$+$sd^{-4}pf^{4}$ configurations using the KSHELL code \cite{Shimizu}.
The quenching factor for the axial-vector coupling constant, $q_A$ = $g_A^{eff}$/$g_A$, is taken to be $q_A$ =1.   
When the phenomenological SDPF-VMU-LS interaction was used for the $sd$-$pf$ cross shell part, the configuration space was limited to $sd^{-2}pf^{2}$ and $q_A$ = 0.775 was adopted \cite{SH2013}. 
The calculated results for the cumulative sum of the $B(GT)$ obtained for EEdf2S and SDPF-VMU-LS as well as the experimental data \cite{Bhat} are shown in Fig. 1(a).
$B(GT)$ is defined by
\begin{equation}
B(GT) = \frac{1}{2J_i +1}\mid <f|| q_A \sum_{i} \vec{\sigma}_i t_{-, i} ||i>\mid^2
\end{equation}
where $J_i$ is the spin of the initial state and $t_{-}|n>$ = $|p>$, and the sum over nucleons i is taken. 
Both the EEdf2S and the SDPF-VMU-LS interactions explain rather well the $B(GT)$ strengths in $^{40}$Ar, though the strength is lower in the low excitation energy region and higher in the high excitation energy region for EEdf2S.
The charged-current reaction cross sections for $^{40}$Ar ($\nu_e$, $e^{-}$) $^{40}$K for the excitations of the 1$^{+}$ and 0$^{+}$ states are compared in Fig. 1(b).
The cross sections obtained by the two interactions are found to be close to each other.
The difference between them is as small as within 5$\%$. 
The cross sections for $^{40}$Ar ($\nu_e$, $e^{-}$) $^{40}$K obtained by the hybrid model in Ref. \cite{SH2013} thus remain almost unchanged when the EEdf2S interaction is used for the GT part.

\section{Neutral-current reactions}
\subsection{Magnetic dipole strength in $^{40}$Ar}

\begin{figure*}[tbh]
\begin{center}
\hspace{-5mm}
\includegraphics[scale=0.44]{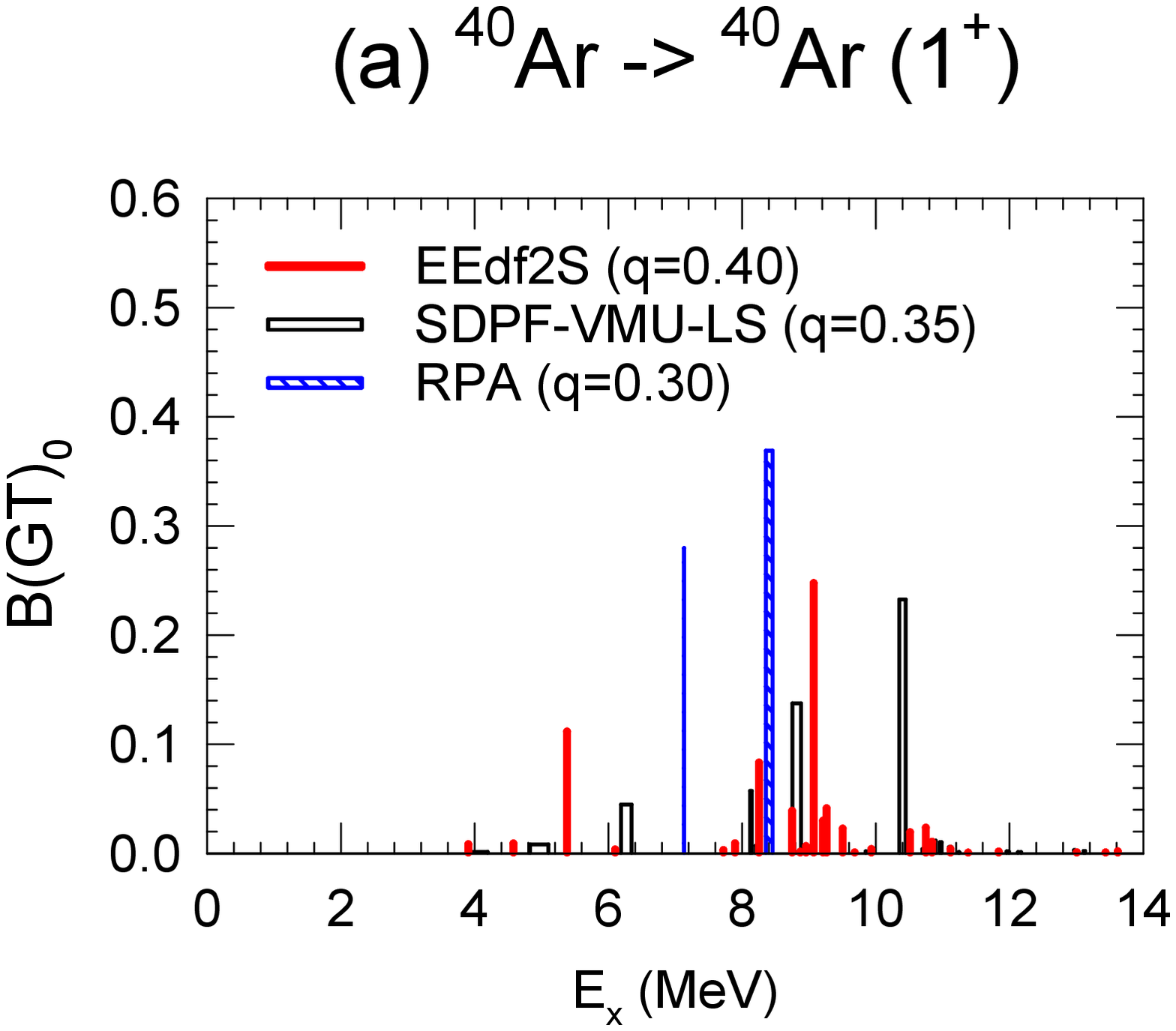}
\hspace*{-10mm}
\includegraphics[scale=0.44]{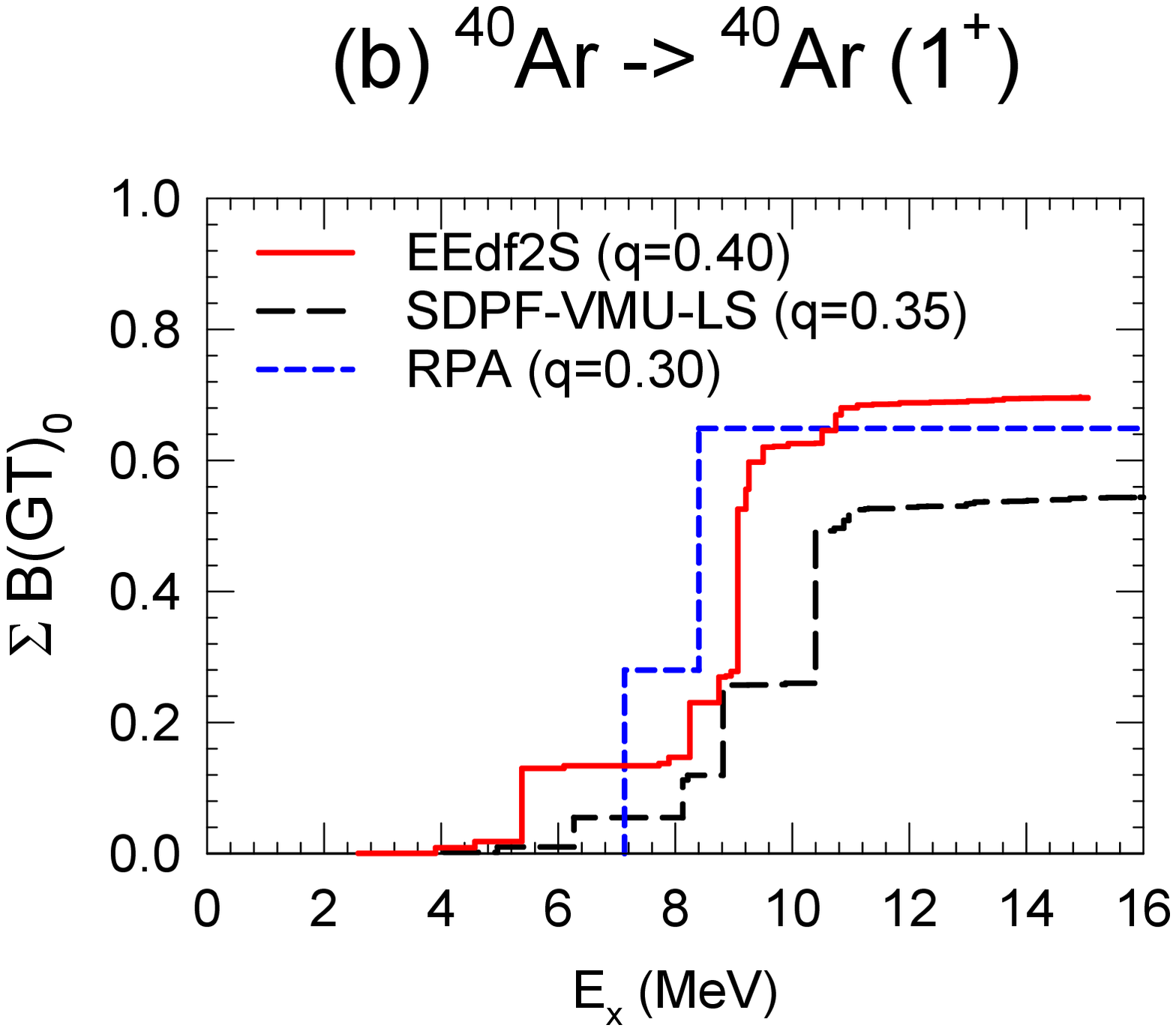}
\vspace{-30mm}
\caption{(Color online)
(a) Red-filled, black-hollow, and blue-shaded histograms denote calculated $B(GT)_0$ values in $^{40}$Ar obtained by shell-model calculations with the EEdf2S, SDPF-VMU-LS as well as by RPA calculations with the SGII with the quenching factors for $g_A$, $q_A$ = 0.40, 0.35 and 0.30, respectively.
(b) The same as in (a) for the cumulative sum of the $B(GT)_0$ up to excitation energies $E_x$.
\label{fig:fig3}}
\end{center}
\end{figure*}

Now we discuss neutral-current reactions on $^{40}$Ar.
Experimental information on the magnetic dipole (M1) strength in $^{40}$Ar is available \cite{Li2006}. The M1 strength was measured by linearly polarized monochromatic $\gamma$ scatterings on $^{40}$Ar in the range of excitation energies, $E_x$ = 7.7-11 MeV. 
One peak of the strength was found at $E_x$ = 9.757 MeV with $B(M1)$ = 0.149$\pm$0.059 $\mu_{N}^2$ \cite{Li2006}.
The $B(M1)$ is defined by
\begin{equation}
B(M1) = \frac{1}{2 J_i +1}\mid <f|| \sqrt{\frac{3}{4\pi}} (g_s\vec{s}_i + g_{\ell}\vec{\ell}_i) \mu_N ||i>\mid^2
\end{equation}
where $g_s$ and $g_{\ell}$ are spin and orbital $g$ factors, respectively, and $\mu_N$ is the nucleon magneton.
The quenching of the $g$-factors is taken into account. 
The isovector (isoscalar) quenching factor for $g_s$ is defined by $q_s^{IV}$ = $g_s^{IV, eff}$/$g_s^{IV}$ ($q_s^{IS}$ = $g_s^{IS, eff}$/$g_s^{IS}$) with $g_s^{IV}$ = -4.70 ($g_s^{IS}$ =0.88).
The isovector orbital $g$ factor is modified by $\delta g_{\ell}^{IV}$ -(0.10-0.15) due to meson-exchange current contributions \cite{Towner,Arima}.
Here, $g_{\ell}$ is taken to be $g_{\ell}$ = 1.15 for proton and $g_{\ell}$ = -0.15 for neutron.

The $B(M1)$ is evaluated by shell-model calculations with the use of the EEdf2S with $sd^{-2}pf^2$+$sd^{-4}pf^4$ configurations.
Calculated results for the quenching $q_s^{IV}$ = $q_s^{IS}$ =0.4 are shown in Fig. 2(a). 
The peaks shown in the figure are for the transitions to the 1$^{+}$ states with isospin $T$=2. 
Note that the main transitions to $T$=1 states in the charged-current case are missing for the neutral-current channel.
The quenching of spin modes in the transitions for $T$=2 $\rightarrow$ $T$=2 could be different from those for $T$=2 $\rightarrow$ $T$=1.
Higher isospin states generally need more configurations to construct the eigenstates of the isospin \cite{Suzuki}.       
The experimental $B(M1)$ strength at $E_x$ = 9.757 MeV is found to be well reproduced with the EEdf2S although the peak energy is shifted towards the lower energy region by 0.69 MeV.
The dominant contributions come from proton d$_{5/2}$ $\rightarrow$ d$_{3/2}$ transition. 
When considering the experimental error bar for $B(M1)$, the quenching factor is obtained to be $q_s^{IV}$ =0.39$\pm$0.04 if $q_s^{IS}$ = $q_s^{IV}$ is assumed.
It is not possible to get a unique $q_s$ from one observed $B(M1)$ value. 
Setting different values for $q_s^{IV}$ and $q_s^{IS}$, for example, $q_s^{IS}$ = 2$q_s^{IV}$, the experimental $B(M1)$ value is also well reproduced for $q_s^{IV}$ =0.35$\pm$0.04.
The calculated $B(M1)$ for the case of $g_{\ell}$=0 are  shown by thin hollow histograms in Fig. 2(a).
The disappearence of the peak at $E_x$ = 2.58 MeV denotes that the 1$^{+}$ state has no spin components and the M1 mode is a pure orbital motion.

Calculated $B(M1)$ obtained for the SDPF-VMU-LS interaction used in Ref. \cite{SH2013} for the $sd^{-2}pf^2$ configurations are shown in Fig. 2(b). 
The quenching factors, $q_s^{IV}$ =0.35 and $q_s^{IS}$ =0.70, and the same $g_{\ell}$ as for EEdf2S are used.
The experimental $B(M1)$ is rather well explained, although the peak is at $E_x$ = 10.40 MeV, which is 0.64 MeV above the experimental energy.   
For $q_s^{IV}$ = $q_s^{IS}$ =0.35, there appear two M1 peaks with similar strength at $E_x$ = 8.82 and 10.40 MeV. 
The former peak has dominant contributions from neutron f$_{7/2}$ $\rightarrow$ f$_{5/2}$ transition. 
This suggests that a higher $q_s^{IS}$ value is favored for SDPF-VMU-LS. 

The $B(M1)$ values are also evaluated by RPA.
A simple configuration, $\pi$d$_{3/2}^{-2}$ $\nu$f$_{7/2}^2$ outside the $^{40}$Ca core is asuumed for the ground state.
The calculated $B(M1)$ obtained with the SGII interaction \cite{SG2} is shown in Fig. 2(b) for $q_s^{IV}$=0.30 and $q_s^{IS}$=0.60. 
Two peaks are obtained at $E_x$ = 7.13 MeV and 8.40 MeV.
Dominant contributions come from proton d$_{5/2}$ $\rightarrow$ d$_{3/2}$ and neutron f$_{7/2}$ $\rightarrow$ f$_{5/2}$ transitions for the state at 7.13 MeV and 8.40 MeV, respectively.
When $q_s^{IV}$ = $q_s^{IS}$ =0.30 is adopted, the first peak with proton components disappears and the second peak with neutron components remains with enhanced strength.
The pure orbital M1 mode is not seen as it is shifted below the ground state in the present calculation. 
The agreement of the calculated $B(M1)$ with the experiment is not so good as the shell-model calculations.

\subsection{Gamow-Teller strength and reaction cross sections for 1$^{+}$}
In this subsection, we discuss GT transitions in the neutral-current channel and reaction cross sections for the 1$^{+}$ multipole.
The GT transition strength, $B(GT)_0$, in the non-charge-exchange channel is defined by
\begin{equation}
B(GT)_0 = \frac{1}{2 J_i +1} \mid <f|| q_A \sum_i \vec{\sigma}_i \tau_{z}^{i} ||i>\mid^2
\end{equation}


\begin{figure*}[tbh]
\includegraphics[scale=0.50]{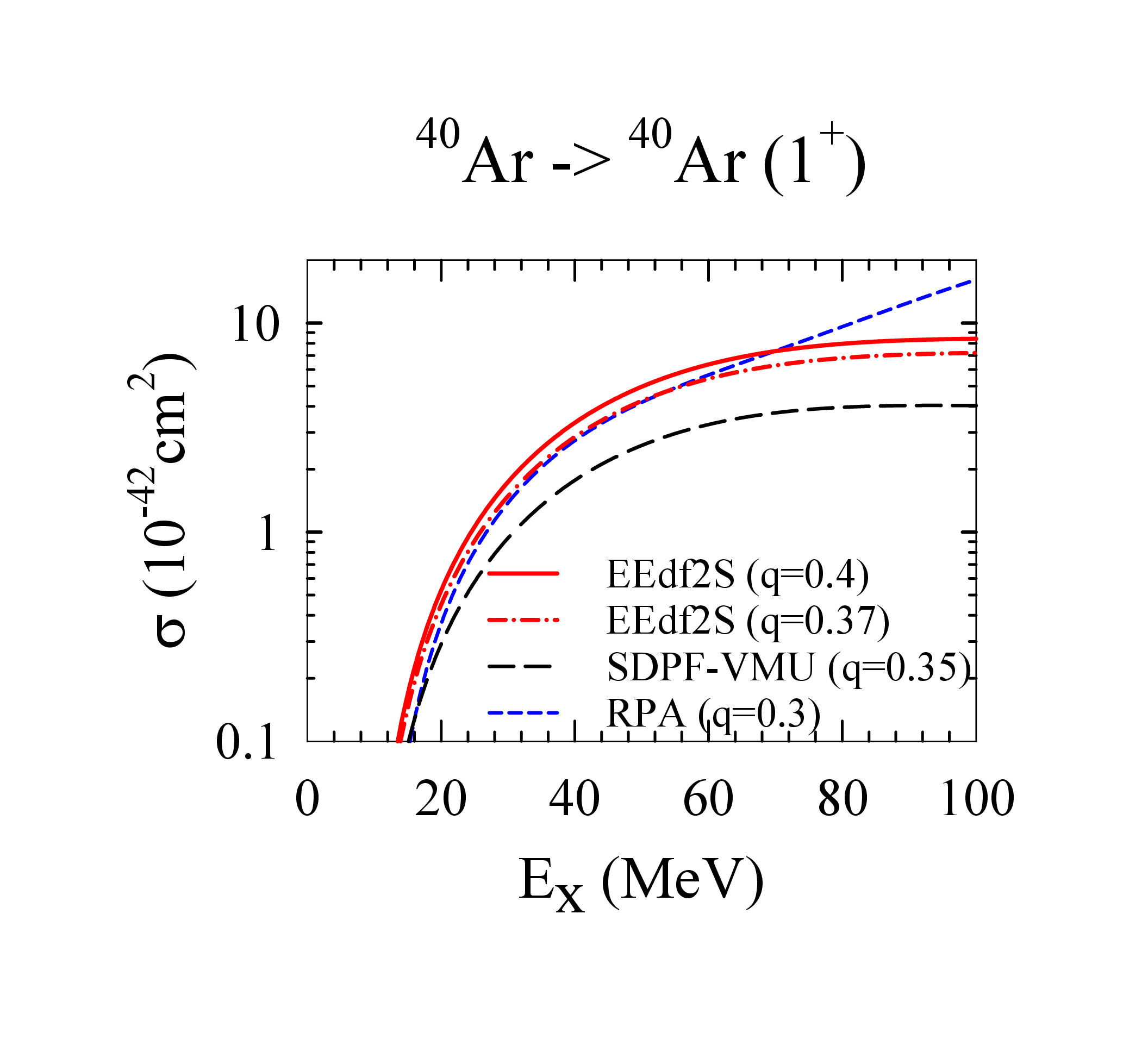}
\vspace{-10mm}
\caption{(Color online)
Calculated cross sections for $^{40}$Ar ($\nu$, $\nu'$) $^{40}$Ar for the 1$^{+}$ multipole, obtained by shell-model calculations with the EEdf2S for $q_A$ = 0.4 and 0.37 and the SDPF-VMU-LS for $q_A$ =0.35.
Results for RPA calculations with $q_A$ =0.3 are also shown.  
\label{fig:fig4}}
\end{figure*}

Calculated $B(GT)_0$ obtained by shell-model calculations with the EEdf2S and SDPF-VMU-LS as well as RPA calculations are shown in Fig. 3(a).
The quenching factor for $g_A$ is taken to be the same as $g_s^{IV}$, that is, $q_A$ = 0.40, 0.35 for EEdf2S and SDPF-VMU-LS, respectively, and $q_A$ =0.30 for RPA with the SGII.
The strength is most spread for the EEdf2S with the largest configuration space among the three cases.
The cumulative sums of $B(GT)_0$ are also shown in Fig. 3(b).
The total strength is the largest for the EEdf2S.

Cross sections for the 1$^{+}$ multipole are evaluated for the shell model and RPA. The energy of the strength is shifted so that the main peak of the $B(GT)_0$ with the dominant proton $d$ orbit components becomes equal to the experimental energy, $E_x$ = 9.757 MeV.
The calculated results are shown in Fig. 4 for the EEdf2S with $q_A$ =0.40 and 0.37, SDPF-VMU-LS with $q_A$ =0.35 and RPA with $q_A$ =0.30.
The cross section for the RPA is close to that for the EEdf2S with $q_A$ =0.37 though the GT strength distributions are different; little spreading of the strength for the RPA case.

\subsection{Neutral-current reaction cross sections}

\begin{figure*}[tbh]
\hspace{-12mm}
\includegraphics[scale=0.47]{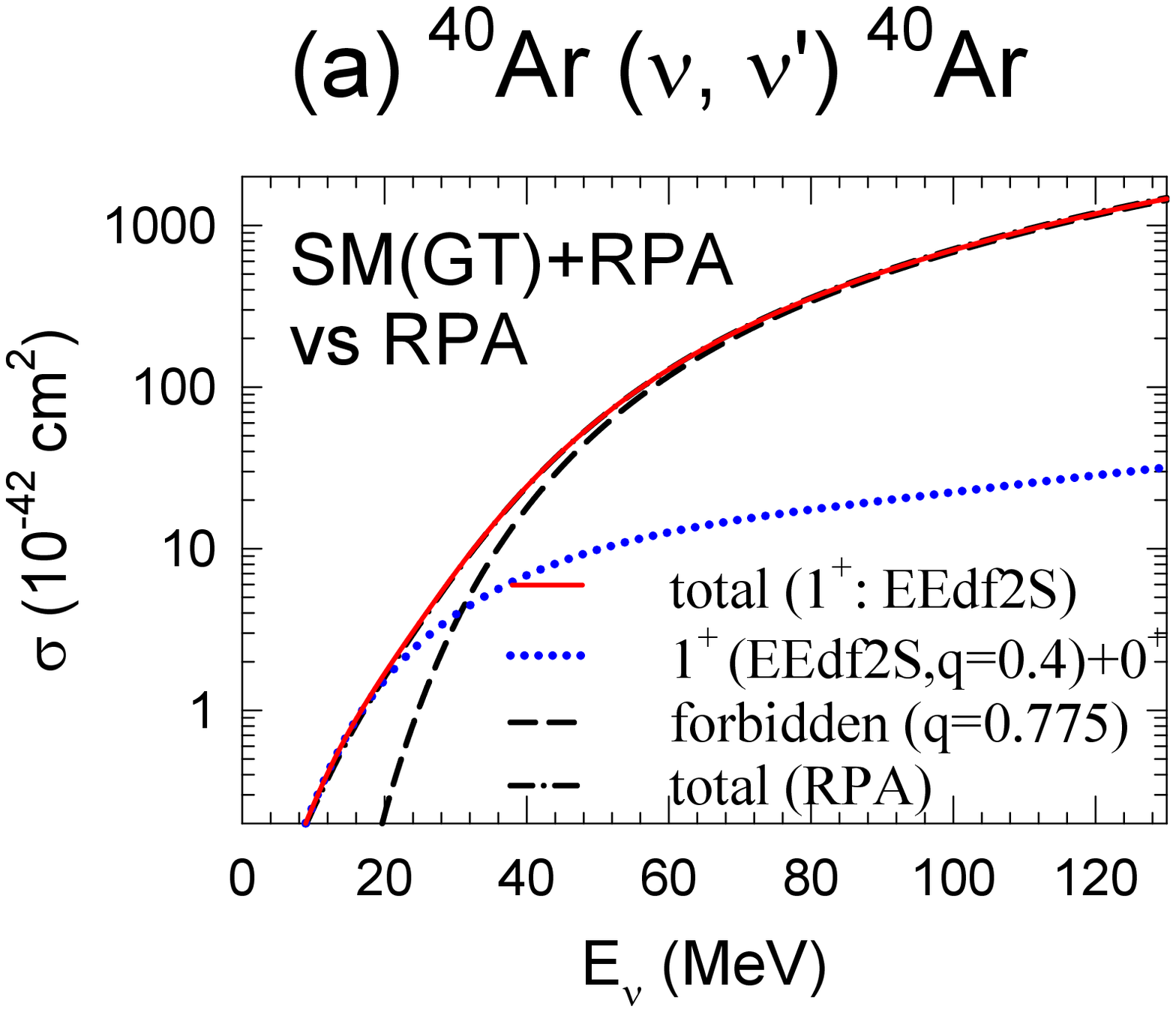}
\hspace{-13mm}
\includegraphics[scale=0.46]{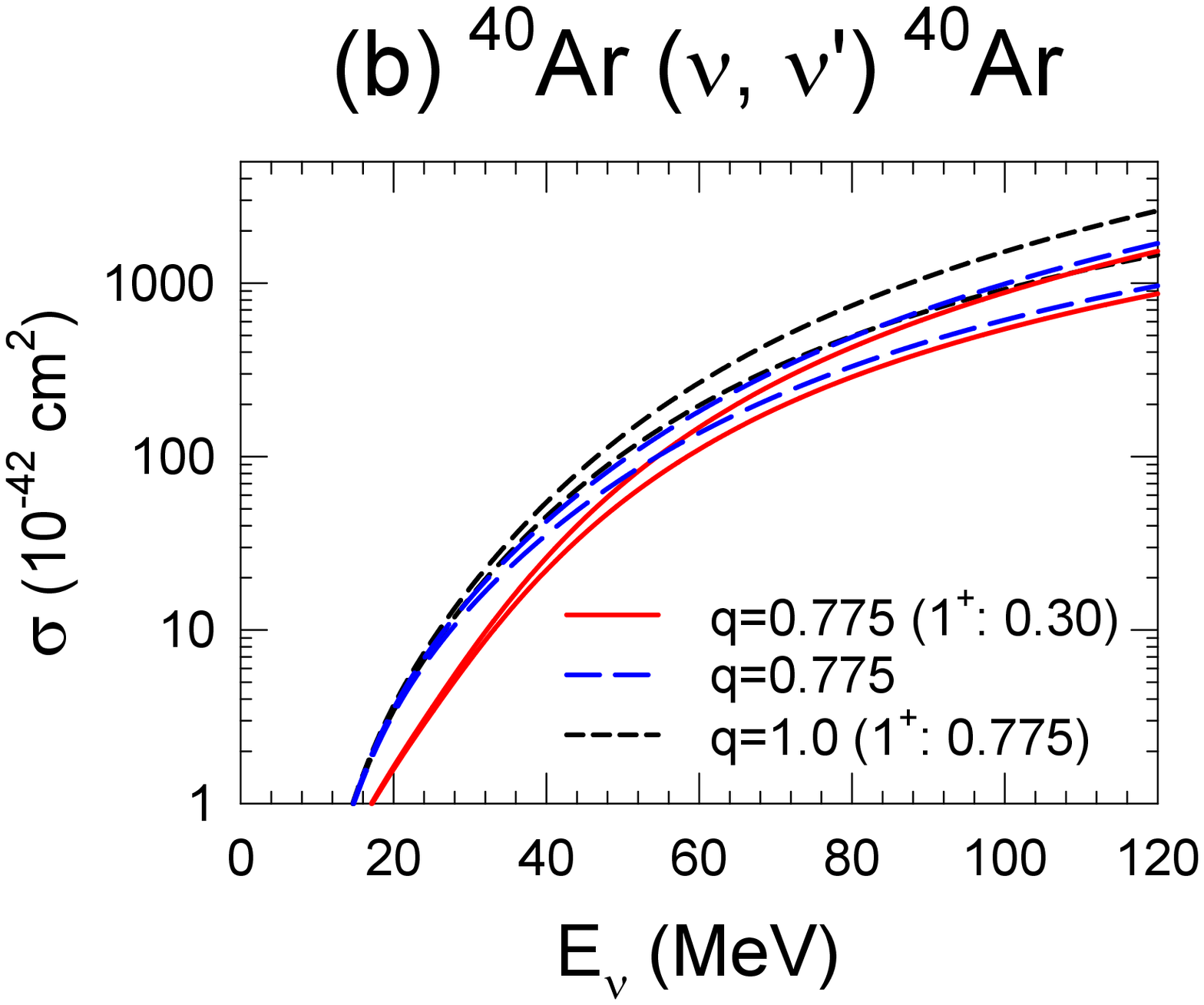}
\vspace{-40mm}
\caption{(Color online)
(a) Calculated neutral-current reaction cross sections obtained by the hybrid model.
The contributions from the 1$^{+}$ multipole are evaluated by the shell model with the EEdf2S, while forbidden transitions are obtained by RPA.
The total cross sections, cross sections for 1$^{+}$ and 0$^{+}$ multipoles and those for the forbidden transitions are denoted by solid, dotted and dashed curves, respectively.
Cross sections evaluated by RPA for all the multipoles are shown by the dash-dotted curve.
(b) Dependence of the total cross sections obtained by RPA on the quenching factor of $g_A$.
Calculated cross sections with $q_A$ =0.30 for 1$^{+}$ and $q_A$ =0.775 for other multipoles are shown by solid curves.
Those with $q_A$ =0.775 for all the multipoles are shown by long-dashed curves, while short-dashed curves denote those with $q_A$ =0.775 for 1$^{+}$ and $q_A$=1.0 for other multipoles.
Larger cross sections denote those for ($\nu$, $\nu'$) reactions, while smaller ones are for ($\bar{\nu}$, $\bar{\nu}'$) reactions. 
\label{fig:fig5}}
\end{figure*}

As the neutrino energy increases, contributions from spin-dipole transitions become important in addition to the GT transitions.
The contributions of multipoles except for 1$^{+}$ are evaluated by RPA calculations.
The sum of the spin-dipole strength, $S^{\lambda}$(SD), energy-weighted sum of the strength, $EWS^{\lambda}$, and averaged energy, $\bar{E}^{\lambda}$, are defined as \cite{Suzuki2018}
\begin{eqnarray}
S^{\lambda} (SD) &=& \sum_{i, \mu} \mid<\lambda \mu, i |S^{\lambda}_{\mu}| 0> \mid^2 \nonumber\\
EWS^{\lambda} &=& \sum_{i, \mu} (E_i -E_0) \mid <i| S^{\lambda}_{\mu}|0>\mid^2 \nonumber\\
 &=& < 0| [S^{\lambda^{\dagger}}, [H, S^{\lambda}]] | 0 > \nonumber\\
\bar{E}^{\lambda} &=& EWS^{\lambda}/S^{\lambda}(SD)
\end{eqnarray}
for the spin-dipole operator
\begin{equation}
S^{\lambda}_{\mu} = [r Y^{1}(\hat{r}) \times \vec{\sigma}]^{\lambda}_{\mu} \tau_z
\end{equation}
The energy-weighted sums for the kinetic energy $K=\sum_{i} \frac{p_{i}^2}{2m}$ with $m$ the nucleon mass, and one-body spin-orbit potential, $V_{LS} = -\xi \sum_{i}\vec{\ell}_{i} \cdot\vec{\sigma}_{i}$, are given as \cite{TS,Suzuki2018}
\begin{eqnarray}
EWS^{\lambda}_{K} &=& \frac{3}{4\pi} h_{\lambda}
\frac{\hbar^2}{2m} A
[1 + \frac{f_{\lambda}}{3A}<0\mid \sum_{i} \vec{\ell}_{i} \cdot\vec{\sigma}_{i} \mid 0>] \nonumber\\
EWS^{\lambda}_{LS} &=& \frac{3}{4\pi} h_{\lambda}
\frac{f_{\lambda}}{3}\xi
<0 \mid \sum_{i} (r_i^2 + g_{\lambda}r_i^2 \vec{\ell}_i \cdot\vec{\sigma}_i) \mid 0> \nonumber\\
\end{eqnarray}
where $h_{\lambda}$ = $2\lambda +1$, $f_{\lambda}$ = 2, 1 and -1 for $\lambda^{\pi}$ = 0$^{-}$, 1$^{-}$ and 2$^{-}$, respectively, and $g_{\lambda}$ = 1 for $\lambda^{\pi}$ = 0$^{-}$, 1$^{-}$ and $g_{\lambda}$ = -7/5 for $\lambda^{\pi}$ = 2$^{-}$.

For $^{40}$Ar, as the term $<0 \mid \sum_{i} \vec{\sigma}_{i}\cdot \vec{\ell}_{i} \mid 0>$ does not vanish and has positive value for d$_{3/2}^{-2}$ f$_{7/2}^2$ configuration, 
$EWS^{\lambda}_{K}/(2\lambda +1)$ decreases as the value of $\lambda$ increases.  
Note that $EWS^{\lambda =2}_{LS}$ is also reduced by the spin-orbit potential.
As the sum of the strength is roughly proportional to $2\lambda$+1, the averaged energy is expected to follow the order: $\bar{E}^{2}$ $<$ $\bar{E}^{1}$ $<$ $\bar{E}^{0}$.  
For the Hamiltonian of the sum of kinetic energy and one-body spin-orbit potential, $S^{\lambda}(SD)$ = 12.53, 36.36 and 56.47 fm$^2$, $EWS^{\lambda}_{K+LS}$ = 333.2, 796.7 and 814.1 MeV$\cdot$fm$^2$, and $\bar{E}^{\lambda} $= 26.58, 21.92 and 14.42 MeV for 0$^{-}$, 1$^{-}$ and 2$^{-}$, respectively.   
The order of the averaged energies is $\bar{E}^{2}$ $<$ $\bar{E}^{1}$ $<$ $\bar{E}^{0}$ as expected. 
The kinetic energy and one-body spin-orbit interaction in the Hamiltonian lead to the splitting of the spin-dipole strength.
Two-body spin-dependent interactions further affect the distribution of the strength.   
When spin-dependent interactions in SGII are added, $S^{\lambda}(SD)$ become 14.87, 57.02 and 52.04 fm$^2$, $EWS^{\lambda}$ are enhanced to 369.6, 1241.0 and 817.4 MeV$\cdot$fm$^2$, and $\bar{E}^{\lambda}$ are found to be 24.85, 21.76 and 15.71 MeV for 0$^{-}$, 1$^{-}$ and 2$^{-}$, respectively.
The sum of the strength and the EWS for 1$^{-}$ are found to be greatly enhanced by the spin-dependent interactions, while the order of $\bar{E}^{\lambda}$ remains the same.   

Reaction cross sections for $^{40}$Ar ($\nu$, $\nu'$) $^{40}$Ar for multipoles other than 1$^{+}$ are obtained by RPA calculations with SGII including up to $J^{\pi}$ = 4$^{\pm}$.
The quenching for $g_A$ is taken to be $q_A$ =0.775 \cite{SH2013}.
Calculated cross sections for 0$^{+}$ and 1$^{+}$ multipoles, for forbidden transitions as well as for the total contributions are shown in Fig. 5(a).
The 1$^{+}$ multipole part is obtained by shell model with the EEdf2S with the quenching factor $q_A$ =0.40.
Contributions from forbidden transitions become important at $E_{\nu} >$ 30 MeV.
As the sum of the strength is largest for 1$^{-}$, the contributions from 1$^{-}$ multipole become most important at $E_{\nu} >$ 40 MeV, while those from 2$^{-}$ multipole are more important at $E_{\nu} <$ 40 MeV due to its lower averaged energy compared with 1$^{-}$.
We notice that when the 1$^{+}$ multipole part is evaluated by RPA with $q_A$ =0.30 instead of the shell model, the total cross section remains almost unchanged.   
In the present work, the quenching factor for $g_A$ in the 1$^{+}$ multipole is constrained by the experimental $B(M1)$ data. 

Calculated cross sections are sensitive to the choice of the quenching factor for $g_A$.
Dependence of the total cross sections obtained by RPA calculations on the quenching factor of $g_A$ is shown in Fig. 5(b) for both neutrino and anti-neutrino scatterings. 
When the quenching for the 1$^{+}$ multipole is taken to be the same as in the charged-current reaction, that is, $q_A$ =0.775 instead of $q_A$ =0.30, the cross sections are enhanced more than twice at $E_{\nu} <$ 40 MeV.
When $q_A$ =1 is adopted instead of $q_A$ =0.775 for the forbidden transitions as usually done in many RPA calculations, the cross sections are enhanced also at higher $E_{\nu}$ and become close to those in Refs. \cite{Jacho,Bottela}. 
It is, thus, important to determine $q_A$ carefully.

\section{Summary}
$\nu$-induced neutral- and charged-current reaction cross sections on $^{40}$Ar are studied by a hybrid model, where the GT transitions and forbidden transitions are treated by the shell model and RPA, respectively.
An effective interaction in $sd$-$pf$ shell constructed by the EKK method \cite{TO2017,TO2020}, referred to as EEdf2S, is used to evaluate $B(GT)$, $B(M1)$ and reaction cross sections for the 1$^{+}$ multipole with a wide configuration space, $sd^{-2}pf^2$+$sd^{-4}pf^4$.
Calculated $B(GT)$ in $^{40}$Ar reproduces rather well the experimental data with $q_A$ =$g_A^{eff}$/$g_A$ = 1.
The calculated cross sections for the $^{40}$Ar ($\nu_e$, $e^{-}$) $^{40}$K exciting 1$^{+}$ states are found to be close to those obtained by the SDPF-VMU-LS interaction in Ref. \cite{SH2013}.

Then, EEdf2S is used to study $B(M1)$ in $^{40}$Ar and neutral-current reaction $^{40}$Ar ($\nu$, $\nu'$) $^{40}$K.
A considerable quenching for the spin g-factor, $q_{s}^{IV}$ $\approx$0.4 is found to reproduce the experimental $B(M1)$ data \cite{Li2006}.
The $B(GT)_0$ in the non-charge-exchange channel and neutral-current reaction cross sections are evaluated, and compared with those obtained by the SDPF-VMU-LS interaction and RPA calculations.
Cross sections for the shell model with the EEdf2S and RPA, where the quenching factors for $g_A$ are determined to be consistent with the experimental $B(M1)$ data, are found to be close to each other in spite of the difference in the GT distributions. 

Cross sections induced by forbidden transitions are obtained by RPA using the SGII interaction.
Contributions from forbidden transitions become important at high neutrino energy, $E_{\nu} >$ 30 MeV.
The total cross sections are obtained, and their sensitivity to the choice of the quenching factor of $g_A$ is indicated.
In the present work, the quenching of spin modes is determined by the $B(M1)$ data available in the range of the excitation energy at $E_x$ = 7.7-11 MeV \cite{Li2006}.
More experimental data for $B(M1)$ at lower excitation energies is required to determine more precise values for the quenching factors.
Further expansion of shell-model configuration space is also an important issue in future.

\acknowledgements
T. S. acknowledges support in part by the Grant-in-Aid for Scientific Research
under Grant Nos. JP19K03855, JP20K03988 of the JSPS.
N. S. acknowledges ``Program for Promoting Researches on the Supercomputer Fugaku'' (JPMXP1020200105) and MCRP program, University of Tsukuba (wo22i022).

\end{document}